\documentstyle[epsf]{article}

\font\tenrm=cmr10
\font\tenit=cmti10
\font\elevenbf=cmbx10 scaled\magstep 1
\font\elevenrm=cmr10 scaled\magstep 1
\font\elevenit=cmti10 scaled\magstep 1

\renewenvironment{thebibliography}[1]
 { \elevenrm
   \begin{list}{\arabic{enumi}.}
    {\usecounter{enumi} \setlength{\parsep}{0pt}
     \setlength{\itemsep}{3pt} \settowidth{\labelwidth}{#1.}
     \sloppy
    }}{\end{list}}

\begin{document}
\baselineskip=13pt

\rightline {\bf UG-FT-36/94}

\begin{center}
\vglue 1.5cm
{
 {\elevenbf        \vglue 10pt
               Model-independent determination of
            ${\bf Z'}$ couplings at LEP 200
               \footnote{
\tenrm
\baselineskip=11pt
To appear in Proceedings of the Zeuthen Workshop on Elementary Particle
Theory "{\tenit Physics at LEP 200 and Beyond}", Teupitz,
Germany, April 10-15, 1994\\}
\vglue 5pt}

\vglue 1.5cm
{\tenrm F. DEL AGUILA \\}
\baselineskip=13pt
\vglue 0.2cm
{\elevenit Departamento de F\'\i sica Te\'orica y del Cosmos,
Universidad de Granada \\}
\baselineskip=12pt
{\elevenit Granada, 18071, Spain \\}
}

\vglue 1.5cm
{\tenrm ABSTRACT}

\end{center}

\vglue 0.3cm
{\rightskip=3pc
 \leftskip=3pc
 \tenrm\baselineskip=12pt
 \noindent
All couplings of a new heavy gauge boson to ordinary
fermions can be determined in a model-independent way
at a large $e^+e^-$ collider for large enough statistics.
No such determination is possible at LEP 200, however,
because its design integrated luminosity is too low for
existing and forthcoming limits on $Z'$ interactions.
At any rate, it should be possible to distinguish between
models for $Z'$ masses $< 1\ TeV$, or to set
limits of this order on $Z'$ masses for definite models.
These are (preliminary) $1\sigma $ limits. We neglect
systematic errors and assume an integrated luminosity
of $0.5\ fb^{-1}$.}

\vglue 1.cm
{\centerline{\elevenbf 1. Introduction}}
\vglue 0.5cm

\elevenrm
There is no compulsary reason for expecting a new heavy
$Z'$ boson, but there is neither a reason for excluding
its existence. At present there are only bounds on its
effects [1-6]. The question here is: Which information
on an extra gauge boson can LEP 200 provide ? At a large
$e^+e^-$ collider if the $Z'$ effects are sizeable all
$Z'$ couplings to ordinary fermions can be determined in
a model-independent way [7,8]. However, the expected
luminosity at LEP 200 is too low to allow for such a
model-independent analysis. The typical value of the $Z'$
mass limit, or of the discovered $Z'$ mass, at TEVATRON
when LEP 200 enters in operation is $\sim 500\ GeV$ [9].
For $Z'$ masses $< 1\ TeV$ it is possible to
distinguish between definite models at LEP 200;
for specific models
it is also possible to set limits on $Z'$ masses $\sim 1\
TeV$. Our analysis is preliminary. In what follows we
review present and forthcoming $Z'$ limits (1.1) and the
parametrization of $Z'$ effects in two-fermion final state
observables at $e^+e^-$ (1.2). Our numerical estimates
for $Z'$ physics at LEP 200 are given in Section 2.

\vskip 0.5cm
{\centerline
{\elevenit 1.1 Present and forthcoming $Z'$ limits}}
\vskip 0.5cm

Present limits on $Z'$ masses, $M_{Z'}$, come from the
absence of an excess of $e^+e^- (\mu ^+\mu ^-)$ pairs
with a large invariant mass at TEVATRON: {\elevenit
direct limits}  [1,2]; and from the observation of no effect
beyond the minimal standard model in precise electroweak
data, mainly at LEP: {\elevenit indirect limits} [3-6].
These
limits for popular models, $\chi , \psi , \eta , LR$,
are given in Table 1 [9,10].
TEVATRON can provide limits on
$M_{Z'}\sim 500(600)\ GeV$
for an integrated luminosity
of $25 (100)\ pb^{-1}$. LEP
allows for a precise measurement of the $Z$ current,
providing a stringent limit on the $Z'Z_1$ mixing,
$s_3\equiv sin\theta _3 < 0.01$ [3,6]. In models where this
mixing is fixed by the $Z'$ mass, the limit on $s_3$
translates into stringent $M_{Z'}$ limits: {\elevenit
constrained indirect limits}. When $M_{Z'}$ and $s_3$ are independent
the $Z'$ mass limits are modest: {\elevenit
unconstrained indirect limits}. The indirect limits on
$M_{Z'}$ depend little on the top quark mass, $m_t$,
except for top masses near the $m_t$ upper bound [5].

\vglue 0.2cm
\begin{center}
\begin{tabular}{|c|ccccc|}
\hline
\multicolumn{1}{|c}{} & $\begin{array}{c}
{\rm direct} \\ ({\rm present}) \\ \end{array}$
& $\begin{array}{c}
{\rm direct} \\ (\int Ldt=25pb^{-1}) \\ \end{array}$
& $\begin{array}{c}
{\rm direct} \\ (\int Ldt=100pb^{-1}) \\ \end{array}$
& $\begin{array}{c}
{\rm indirect} \\ {\rm (constrained)} \\ \end{array}$ &
\multicolumn{1}{c|}{$\begin{array}{c}
{\rm indirect} \\ {\rm (unconstrained)} \\ \end{array}$} \\
\hline
$ \chi $ & $340$ & $470$ & $620$ & $670$ & $380$ \\
$ \psi $ & $320$ & $450$ & $600$ & $200$ & $200$ \\
$ \eta $ & $340$ & $460$ & $610$ & $440$ & $210$ \\
$ LR $ & $310$ & $510$ & $660$ & $990$ & $430$ \\
\hline
\end{tabular}
\end{center}
\vglue 0.1cm

\vglue 0.1cm
\rightskip=3pc
\leftskip=3pc
{\tenrm \baselineskip=12pt
\noindent
Table 1. $Z'$ mass limits (in $GeV$).}
\vglue 0.1cm

\rightskip=0pc
\leftskip=0pc
In conclusion we will assume for our estimates at LEP 200
that a new $Z'$ may exist with a mass $> 500\ GeV$.

\vskip 0.5cm
{\centerline
{\elevenit 1.2 Parametrization of $Z'$ effects in two-fermion
final state observables at $e^+e^-$}}
\vskip 0.5cm

The effects of a new $Z'$ can be observed in the two-fermion
channels at LEP 200. The relevant lagrangian reads:

\begin{equation}
-{\cal L}_{NC}=
e \sum _{i} q_i \bar{\psi}_i \gamma ^{\mu} \psi _i A_{\mu}
+ \frac {g}{2c_W} \sum _{i} \bar{\psi}_i \gamma ^{\mu}
(v^i - a^i\gamma _5) \psi _i Z_{1\mu}
+ \frac {g}{2c_W} \sum _{i} \bar{\psi}_i \gamma ^{\mu}
(v'^i - a'^i\gamma _5) \psi _i Z_{2\mu},
\end{equation}

\noindent
where the last piece describes the new $Z'$ interaction
and the first two pieces define the standard model neutral
lagrangian [1]. (We assume family universality and that the
$SU(2)_L$ charges $T_i$ commute with the new charge $Q':\
[Q', T_i]=0$.) We neglect the $Z'Z_1$ mixing angle, for
it is bounded, $s_3 < 0.01$, and its effects are
unobservable at LEP 200. Hence, we identify $Z_1$ and
$Z_2$ with the observed $Z$ and the new $Z'$ bosons,
respectively. The three gauge bosons, $\gamma , Z, Z'$,
are far off-shell for a center of mass energy of
$200\ GeV$. The expected statistics allows for working
at tree level. Radiative effects due to real $Z$ production
are important but they can be taken away with a cut on the
maximum photon energy [11]. We can neglect them in a first
estimate of the $Z'$ potential of LEP 200. The main $Z'$
effects result from the interference of the $Z'$ amplitude,

\begin{equation}
{\cal A}^f(Z') = \frac {1}{s-M_{Z'}^2+i\Gamma _{Z'}M_{Z'}}
(\frac {g}{2c_W})^2 \bar u_e \gamma ^{\mu}
(v'^e-a'^e\gamma _5) u_e \bar u_f \gamma _{\mu}
(v'^f-a'^f\gamma _5) u_f,
\end{equation}

\noindent
with the $\gamma , Z$ amplitudes. Neglecting the $Z'$
width and for a fixed center of mass energy $\sqrt s\ (=
200\ GeV)$, the $Z'$ amplitude and the corresponding
$Z'$ effects in $e^+e^-\rightarrow f\bar f$, with $f$
any ordinary fermion, depend on five couplings [7]:

\begin{equation}
\epsilon _A = \frac {a'^{e 2}}{c_W^2s_W^2}
\frac {s}{M_{Z'}^2-s}, \
P_V^e = \frac {v'^e}{a'^e},
\ P_L^q = \frac {v'^q+a'^q}
{2a'^e},
\ P_R^{u,d} = \frac {v'^{u,d}-a'^{u,d}}
{v'^q+a'^q}.
\end{equation}

At a large $e^+e^-$ collider not only the pairs of
charged leptons but the bottom and charm final states
can be tagged. As a matter of fact, using the event shape,
lepton tagging and/or vertex reconstruction LEP analyses
have measured the total cross section and
the forward-backward asymmetry of bottom and charm pairs
[12]. Thus, the set of observables at LEP 200
includes the total cross section for charged leptons
$\sigma ^l$
and their forward-backward asymmetry
$A^l_{FB}$ ($l$ stands for $e, \mu , \tau $), as well as for bottom and charm
quarks
$R^{b,c}\equiv \frac {\sigma ^{b,c}}{\sigma ^l},
A^{b,c}_{FB}$,
and the total cross section for hadrons
$R^{had}\equiv \frac {\sigma ^{had}}{\sigma ^l}$.
(We assume only $s$-channel exchange for electrons,
implying $\sigma ^e \sim \sigma ^{\mu} \sim
\sigma ^{\tau}$. We do not consider tau polarization
asymmetries.)
If beam polarization is available, the corresponding
left-right asymmetries, $A^{l,b,c,had}_{LR},
A^{l,b,c}_{LR,FB}$,
double the number of
observables to be measured.

$\sigma ^l$ and $A^l_{FB}$ have small linear terms
in $P^e_V$, implying that unpolarized
lepton probes only determine $\epsilon _A$ and $|P_V^e|$.
Whereas the polarized probes $A^l_{LR}$ and
$A^l_{LR,FB}$ are very sensitive
to $P_V^e$, including its sign [7]. One combination of
$P^q_L, P^{u,d}_R$ is fixed by $R^{had}$.
A second combination is
determined by $A^{had}_{LR}$ if polarization is available.
In any case, flavour
tagging is necessary for a complete determination of the
$Z'$ couplings to quarks. $R^{b,c}$ and $A^{b,c}_{FB}$
allow for this
determination even without electron beam polarization.

In Ref. [7] it is
shown that a complete (model-independent) determination
of the $Z'$ couplings to ordinary fermions is possible
for a $Z'$ mass $\sim 1\ TeV$ at NLC. As we show below
such a determination is not possible at LEP 200.

\vglue 1.cm
{\centerline{\elevenbf 2. LEP 200 results}}
\vglue 0.5cm

In Table 2 we gather the expected values
of two-fermion final state observables at LEP 200 for
different models. We assume an integrated luminosity
of $0.5\ fb^{-1}$. Errors are only statistical. The
first four columns correspond to models $\chi , \psi ,
\eta , LR$ and a $Z'$ mass equal to $500\ GeV$. The
fifth column gives the values of the same observables
for the standard model.

\begin{center}
\begin{tabular}{|c|ccccc|}
\hline
\multicolumn{1}{|c}{} & $\chi$
& $\psi$
& $\eta$
& $LR$ &
\multicolumn{1}{c|}{$SM$} \\
\hline
$\sigma ^lL_{int}$ & $1390\pm 40$ & $1480\pm 40$ &
$1460\pm 40$ & $1460\pm 40$ & $1500\pm 40$ \\
$A^l_{FB}$ & $0.564\pm 0.020$ & $0.542\pm 0.022$
& $0.561\pm 0.022$ & $0.526\pm 0.022$ & $0.557\pm 0.021$ \\
$R ^{had}$ & $7.37\pm 0.20$ & $6.65\pm 0.19$
& $6.86\pm 0.19$ & $6.61\pm 0.19$ & $6.60\pm 0.18$ \\
 &  &  &  &  &  \\
$R ^c$ & $1.84\pm 0.06$ & $1.78\pm 0.06$
& $1.78\pm 0.06$ & $1.70\pm 0.06$ & $1.72\pm 0.06$ \\
$R ^b$ & $1.23\pm 0.04$ & $1.03\pm 0.04$
& $1.10\pm 0.04$ & $1.07\pm 0.04$ & $1.05\pm 0.04$ \\
$A^c_{FB}$ & $0.657\pm 0.015$ & $0.680\pm 0.014$
& $0.667\pm 0.015$ & $0.642\pm 0.015$ & $0.670\pm 0.015$ \\
$A^b_{FB}$ & $0.585\pm 0.020$ & $0.603\pm 0.020$
& $0.621\pm 0.020$ & $0.566\pm 0.020$ & $0.604\pm 0.020$ \\
 &  &  &  &  &  \\
$A^l_{LR}$ & $0.038\pm 0.019$ & $0.086\pm 0.018$
& $0.099\pm 0.018$ & $0.099\pm 0.018$ & $0.086\pm 0.018$ \\
$A^l_{LR,FB}$ & $0.028\pm 0.019$ & $0.064\pm 0.018$
& $0.075\pm 0.018$ & $0.074\pm 0.018$ & $0.064\pm 0.018$ \\
$A^{had}_{LR}$ & $0.484\pm 0.006$ & $0.488\pm 0.006$
& $0.479\pm 0.006$ & $0.561\pm 0.006$ & $0.486\pm 0.006$ \\
 &  &  &  &  &  \\
$A^c_{LR}$ & $0.363\pm 0.013$ & $0.378\pm 0.013$
& $0.356\pm 0.013$ & $0.453\pm 0.013$ & $0.384\pm 0.013$ \\
$A^b_{LR}$ & $0.604\pm 0.014$ & $0.617\pm 0.014$
& $0.612\pm 0.014$ & $0.677\pm 0.013$ & $0.598\pm 0.014$ \\
$A^c_{LR,FB}$ & $0.187\pm 0.014$ & $0.220\pm 0.013$
& $0.190\pm 0.014$ & $0.245\pm 0.014$ & $0.219\pm 0.014$ \\
$A^b_{LR,FB}$ & $0.489\pm 0.015$ & $0.533\pm 0.015$
& $0.523\pm 0.015$ & $0.572\pm 0.015$ & $0.532\pm 0.015$ \\
\hline
\end{tabular}
\end{center}
\vglue 0.1cm

\vglue 0.1cm
\rightskip=3pc
\leftskip=3pc
{\tenrm\baselineskip=12pt
\noindent
Table 2. Values and statistical errors for two-fermion
final state observables at LEP 200
($\sqrt s = 200\ GeV, \int Ldt = 0.5\ fb^{-1}$) for models
$\chi , \psi , \eta , LR\ (M_{Z'} = 500\ GeV)$ and for the
standard model.}
\vglue 0.1cm

\rightskip=0pc
\leftskip=0pc
In order to decide on the $Z'$ potential of LEP 200
we perform a $\chi ^2$ fit to the observables  in Table 2
for models $\chi , \psi , \eta , LR$.
$l$ stands for $e, \mu , \tau $ and counts three times.
The results are given in Table 3. The errors in
parentheses are for unpolarized electron beams. In both
cases the errors are so large that no model-independent
determination of the $Z'$ couplings to ordinary fermions
is possible.

\vglue 0.2cm
\begin{center}
\begin{tabular}{|c|cccc|}
\hline
\multicolumn{1}{|c}{} & $\chi$
& $\psi$ & $\eta$ &
\multicolumn{1}{c|}{$LR$} \\
\hline
$ \epsilon _A $ & $0.041\pm 0.013(0.048)$
& $0.069\pm 0.043(0.076)$ & $0.007\pm 0.009(0.030)$
& $0.145\pm 0.041(0.046)$ \\
& & & & \\
$ P^e_V $ & $2\pm 0.4(1.2)$ & $0\pm 0.2(10.6)$
& $-3\pm 2.2(6.7)$ & $-0.148\pm 0.1(0.4)$ \\
$ P^q_L $ & $-0.5\pm 0.2(0.5)$ & $0.5\pm 0.5(2.0)$
& $2\pm 1.6(5.2)$ & $-0.142\pm 0.2(0.3)$ \\
$ P^u_R $ & $-1\pm 0.9(1.1)$ & $-1\pm 0.7(12.5)$
& $-1\pm 0.7(1.5)$ & $-6.04\pm 7.5(13.9)$ \\
$ P^d_R $ & $3\pm 1.2(2.7)$ & $-1\pm 1.1(26.4)$
& $0.5\pm 0.4(2.5)$ & $8.04\pm 10.2(17.8)$ \\
\hline
\end{tabular}
\end{center}
\vglue 0.1cm

\vglue 0.1cm
\rightskip=3pc
\leftskip=3pc
{\tenrm \baselineskip=12pt
\noindent
Table 3. Values of couplings in Eq. (3) and $1\sigma $
statistical errors at LEP 200 for popular models.
Errors in parentheses
correspond to unpolarized electron beams.}
\vglue 0.1cm

\rightskip=0pc
\leftskip=0pc
However, we can distinguish between
specific models.
Let us consider the class of $E_6$ models parametrized
by $\beta $ in Table 4. Performing a $\chi ^2$ fit to
the observables in Table 2 for models $\chi , \psi , \eta $
with $M_{Z'} = 500\ GeV$, we obtain

\begin{equation}
\beta = 0.0 \pm 0.10(0.27), 1.57 \pm 0.10(0.28),
-0.91 \pm 0.11(0.13),
\end{equation}

\noindent
respectively. Errors in parentheses correspond to unpolarized electron
beams.

\vglue 0.2cm
\begin{center}
\begin{tabular}{|c|cc|}
\hline
\multicolumn{1}{|c}{} & $\frac {g}{e}v'^i$ &
\multicolumn{1}{c|}{$\frac {g}{e}a'^i$} \\
\hline
$ e $
& $\sqrt {\frac {2}{3}} cos \beta $ &
$ \frac {1}{\sqrt 6} cos \beta + \frac {1}{3}
\sqrt {\frac {5}{2}} sin \beta $ \\
$ u $ & $ 0 $ &
$ -\frac {1}{\sqrt 6} cos \beta + \frac {1}{3}
\sqrt {\frac {5}{2}} sin \beta $ \\
$ d $ & $ -\sqrt {\frac {2}{3}} cos \beta $ &
$ \frac {1}{\sqrt 6} cos \beta + \frac {1}{3}
\sqrt {\frac {5}{2}} sin \beta $ \\
\hline
\end{tabular}
\end{center}
\vglue 0.1cm

\vglue 0.1cm
\rightskip=3pc
\leftskip=3pc
{\tenrm \baselineskip=12pt
\noindent
Table 4. Vector and axial $Z'$ couplings in Eq. (1) for
a class of $E_6$ models.
$\beta = 0, \frac {\pi}{2}, -arc tan \sqrt {\frac {5}{3}}$
for models $\chi , \psi , \eta $, respectively.}
\vglue 0.1cm

\rightskip=0pc
\leftskip=0pc
For definite models we can also determine the $Z'$ mass.
A $\chi ^2$ fit to the
observables in Table 2 gives

\begin{equation}
M_{Z'} = 500\pm 19(25), 83(102), 43(72), 14(50),
\end{equation}

\noindent
for models $\chi , \psi , \eta , LR$, respectively.
In the Figure we show for the standard model observables
in Table 2 the $1 \sigma \ M_{Z'}$ limits
for the $E_6$ models in Table 4 as a function of $\beta $.
The solid curve corresponds to unpolarized beams.
%
% The figure
%
\begin{figure}
\setlength{\unitlength}{1truecm}
\begin{picture}(13.0,10.0)
\epsfxsize=12cm
%\put(0.,-6.5){\epsfbox{mzpsm.ps}}
\end{picture}
\caption{}
\end{figure}
\vglue 1.cm
{\centerline{\elevenbf 3. Conclusions}}
\vglue 0.5cm

For LEP 200 or for any center of mass energy of the same
order and an integrated luminosity $\sim 0.5\ fb^{-1}$
no model-independent determination of the $Z'$ couplings
to ordinary fermions is possible for $M_{Z'} > 500\ GeV$.
However, we can distinguish between specific models for
$M_{Z'}$
fixed (and $< 1\ TeV$). Limits on $M_{Z'}\sim 1\ TeV$
can also be obtained for definite models. This analysis can
be improved by including systematic errors and
(initial) radiation.

\vglue 0.8cm
{\elevenbf\noindent Acknowledgements \hfil}
\vglue 0.3cm

We thank M. Cveti\v c, P. Langacker, A. Leike and
S. and T. Riemann for discussions and
the organizers of the workshop for their hospitality.
This work was partially supported by the European Union
under contract CHRX-CT92-0004 and by CICYT.

\vglue 1.cm

{\elevenbf\noindent References \hfil}
\vglue 0.3cm

\end{document}